\documentclass[aps,prl,twocolumn,superscriptaddress,amsfont,amssymb,amsmath,nofootinbib,showpacs,balancelastpage]{revtex4-1}
\usepackage{graphicx,longtable,natbib,mathrsfs,eucal}

\newcommand{\de}{\mathrm d}

\newcommand{\om}{{\Omega_m}}
\newcommand{\ob}{{\Omega_b}}

\newcommand{\odm}{{\Omega_\mathrm{DM}}}

\newcommand{\ho}{{H_0}}
\newcommand{\rc}{{r_c}}
\newcommand{\lcdm}{$\Lambda$CDM}
\newcommand{\cinf}{{c_\infty}}

\begin{document}

\title{The Power of Cosmic Flexion in Testing Modified Matter and Gravity}

\author{Stefano Camera}
\email{camera@ph.unito.it}
\affiliation{Dipartimenti di Fisica Generale e Teorica, Universit\`a degli Studi di Torino, and INFN, Sezione di Torino, via P. Giuria 1, 10125 Torino, Italy}
\author{Antonaldo Diaferio}
\email{diaferio@ph.unito.it}
\affiliation{Dipartimento di Fisica Generale ``A. Avogadro'', Universit\`a degli Studi di Torino; INFN, Sezione di Torino, via P. Giuria 1, 10125 Torino, Italy, and Harvard-Smithsonian Center for Astrophysics, 60 Garden St., Cambridge, MA, USA}

\date{Received \today; published -- 00, 0000}

\begin{abstract}
Flexion is the weak lensing effect responsible for the weakly skewed and arc-like appearance of lensed galaxies. The flexion signal-to-noise ratio can be an order of magnitude larger than that of shear. For the first time, we show how this makes flexion an invaluable tool for discriminating between alternative cosmological models. We analyse a scalar field model of unified dark matter and dark energy, a brane-world cosmology and two $f(R)$ modified-action theories. We show that these models can be distinguished from \lcdm\ at several standard deviations by measuring the power spectrum of cosmic flexion.
\end{abstract}


\maketitle

\textit{Introduction.---}In the last decades, cosmologists proposed several models alternative to the concordance $\Lambda$ cold dark matter (\lcdm) paradigm. These models attempt to find an agreement at least as good as that of \lcdm\ with current cosmological datasets, such as the temperature anisotropy pattern of the cosmic microwave background radiation \citep{Komatsu:2008hk}, the dynamics of the large-scale structure of the Universe \citep{Riess:2006fw} and the present-day cosmic accelerated expansion \citep{Larson:2010gs}. However, in these theories, crucial topics such as the missing mass in galaxies and galaxy clusters and the current Universe's accelerated expansion are not explained by the usual dark matter and the cosmological constant $\Lambda$. On the contrary, these models mainly rely on either a modification of the law of gravity or the introduction of additional scalar or vector fields in the Universe's content.

The family of alternative models with additional fields, also named modified matter models, includes dynamical dark energy or quintessence \citep{2010deto.book.....A}, but also models which attempt to identify both the dark matter and dark energy effects with the properties of a single ``dark fluid'' \citep{Sapone:2010iz,2010AdAst2010E..78B}. Conversely, the class of modified gravity includes a variety of approaches, which can however be well represented by brane-world cosmologies \citep{Maartens:2010ar} and modified-action theories \citep{DeFelice:2010aj}. Brane worlds describe a four-dimensional ``brane,'' which is our own Universe, embedded into a higher-dimensional spacetime, the ``bulk.'' In this scenario, Einstein's general relativity is still valid, but the higher-dimensional behaviour of gravity induces non-negligible signatures on the Universe's evolution and the growth of cosmic structures on the brane. Finally, modified-action theories directly modify the law of gravity by generalising the Hilbert-Einstein Lagrangian. Among all the possible theories, $f(R)$ gravity, where Ricci's scalar $R$ is replaced by a generic function $f(R)$, is probably the most investigated approach.

In this \textit{Letter}, we choose three models to explore the space of modified matter and gravity theories. Specifically, we consider a model of unified dark matter and dark energy \citep{Bertacca:2008uf}, a phenomenological extension of the well-known DGP brane-world cosmology \citep{Afshordi:2008rd}, and two $f(R)$ models able to pass the Solar system gravity tests \citep{Starobinsky:2007hu,Hu:2007nk}. All of them reproduce the \lcdm\ expansion history, thus representing viable alternatives for the description of the background evolution of the Universe. To be able to discriminate between them and \lcdm\ it is therefore crucial to investigate the r\'egime of cosmological perturbations. This analysis has been carried out using several observables, for instance the power spectrum of density fluctuations and cosmic shear \citep{Jain:2010ka}. However, it is not rare that the predicted signal is very similar to what expected in \lcdm. Here, we show that the degeneracy between models can be lifted by cosmic flexion, namely the flexion correlation function whose signal originates from the large-scale structure of the Universe. We will present parameter forecasts and additional gravitational lensing statistics elsewhere \citep{Camera:flexion}.

\textit{Cosmic Flexion.---}The deflecting gravitational field of the extended large-scale structure of the Universe -- which is simply the Newtonian potential $\Phi$, in general relativity -- is responsible for deflection of light rays emitted by distant sources. This phenomenon is known as weak gravitational lensing. Therefore, photon paths from a galaxy located at $\boldsymbol\theta$ on the sky are deflected by an angle
\begin{equation}
\alpha=\partial\psi,
\end{equation}
where $\partial=\partial_1+i\partial_2$ is the gradient with respect to directions perpendicular to the line of sight and $\psi$ is the projected deflecting potential. Unfortunately, the deflection angle is not observable directly. This is because one does not know the true two-dimensional distribution of the sources on the sky. On the other hand, its gradient, the distortion matrix $\partial_a\partial_b\psi$, is measurable. In particular, the entries of the distortion matrix can be related to the effects of convergence $\kappa$ and (complex) shear $\gamma$ occurring to the source image, i.e.
\begin{equation}
\kappa=\frac{1}{2}\partial\partial^\ast\psi,\quad\gamma=\frac{1}{2}\partial\partial\psi.
\end{equation}

If convergence and shear are effectively constant within a source galaxy image, the galaxy transformation is ${\theta'}_a=A_{ab}\theta_b$, where $A_{ab}=\delta_{ab}-\partial_a\partial_b\psi$ and $a,b=1,2$ label the coordinates on the sky. Flexion arises from the fact that the shear and convergence are actually not constant within the image, it therefore represents local variability in the shear field that expresses itself as second-order distortions in the coordinate transformation between unlensed and lensed images. Thus, by expanding the observed galaxy position $\boldsymbol\theta'$ at the second-order in the deflection angle, it follows that ${\theta'}_a\simeq A_{ab}\theta_b+D_{abc}\theta_b\theta_c/2$ \citep{Goldberg:2004hh}, with $D_{abc}\equiv\partial_cA_{ab}$. As the distortion matrix can be decomposed into the convergence and the shear, it is usual to define a spin-1 and a spin-3 flexion, which read
\begin{equation}
\mathcal F=\frac{1}{2}\partial\partial\partial^\ast\psi,\quad\mathcal G=\frac{1}{2}\partial\partial\partial\psi,\label{eq:flexions}
\end{equation}
respectively. Since measurements of $\mathcal G$ are noisier than $\mathcal F$ \citep{Okura:2006fi}, we will restrict our analysis to $\mathcal F$ only.

To construct the flexion correlation function from large-scale structure, we start from the definition of the projected deflecting potential,
\begin{equation}
\psi(\boldsymbol\theta)=\int\!\!\de\chi\,W(\chi)\Phi(\chi,\boldsymbol\theta),
\end{equation}
where $\de\chi=\de z/H(z)$ is the radial comoving distance, $H(z)$ is the expansion history of the Universe and $W(\chi)$
is the weak lensing selection function \citep{Kaiser:1991qi}. $W(\chi)$ depends on the redshift distribution of the sources $n[\chi(z)]$, normalised such that $\int\!\!\de\chi\,n(\chi)=1$.

In the flat-sky approximation, we expand the flexion in its Fourier modes $\mathcal F(\boldsymbol\ell)$. 
Hence, from the definition of angular power spectrum
\begin{equation}
\langle\mathcal F(\boldsymbol\ell)\mathcal F^\ast(\boldsymbol\ell')\rangle=\left(2\pi\right)^2\delta_D(\boldsymbol\ell-\boldsymbol\ell')C^{\mathcal F}(\ell),
\end{equation}
which is the Fourier transform of the two-dimensional correlation function, and from Eq.~\eqref{eq:flexions}, we finally get \citep{Bacon:2005qr}
\begin{equation}
C^{\mathcal F}(\ell)=\frac{\ell^6}{4}\int\!\!\de\chi\,\frac{W^{2}(\chi)}{\chi^2}P^\Phi\left(\frac{\ell}{\chi},\chi\right).\label{eq:C^F_l}
\end{equation}

\textit{Modified Matter/Gravity Models.---} We now briefly review the three models we use. We refer to them as: UDM for the model of unified dark matter and dark energy; eDGP for the phenomenologically extended DGP brane world; and St and HS for the two $f(R)$ theories of \citep{Starobinsky:2007hu} and \citep{Hu:2007nk}, respectively.

In the class of UDM models we use \citep{Bertacca:2008uf}, a single scalar field with a Born-Infeld kinetic term \citep{Born:1934gh} mimics both dark matter and dark energy. The energy density of the scalar field reads $\rho_\mathrm{UDM}=\rho_\mathrm{DM}+\rho_\Lambda$, where $\rho_\mathrm{DM}\propto a^{-3}$ and $\rho_\Lambda=\mathrm{const.}$, that yields the \lcdm\ Hubble parameter exactly. However, there also is a pressure term $p_\mathrm{UDM}=-\rho_\Lambda$ which leads to a non-negligible speed of sound for the perturbations of the scalar field itself. This is a common feature in modified matter models and it typically causes an integrated Sachs-Wolfe effect incompatible with current observations \citep{Bertacca:2007cv}. To solve this problem, we use the technique outlined in \citep{Bertacca:2008uf}, where the authors construct a UDM model able to reproduce both the correct temperature power spectrum of the cosmic microwave background and the clustering properties of the large-scale structure we see today.

The sound speed is parameterised by its late-time value $\cinf$ (in units of $c=1$) and the growth of cosmic structures strongly depends on it. Indeed, the presence of the sound speed produces an effective Jeans length $\lambda_J$ for the Newtonian potential. Thus, its evolution is no more scale independent. Specifically, the Fourier modes $\Phi_k$ are suppressed on scales $k>1/\lambda_J$ and oscillate around zero. The larger is the value of $\cinf$, the earlier the Newtonian potential starts decreasing (for a fixed scale) or at a greater scale (for a fixed epoch) \citep{Camera:2009uz}. Since the Newtonian potential is responsible for light deflection, weak lensing is a powerful tool to constrain UDM models \citep{Camera:2009uz} and in particular three-dimensional cosmic shear \citep{Camera:2010wm}. However, UDM models with $\cinf\lesssim10^{-3}$ still produce a signal virtually indistinguishable from that of \lcdm.

In the eDGP model \citep{Afshordi:2008rd}, the cross-over length $\rc$, which defines the scale at which higher-dimensional gravitational effects become important, is tuned by a free parameter $\alpha\in[0,1]$. It is strictly related to the graviton propagator. Particularly, $\alpha=0$ and $\alpha=1/2$ reduce to \lcdm\ and standard DGP, respectively. Recently, it has been shown that the eDGP model excellently fits geometrical datasets such as the Hubble diagram of type Ia supernov\ae\ and gamma ray bursts, the scale of baryon acoustic oscillations and the CMB distance indicators \citep{Camera:2011mg}. Therefore, it is crucial to test this model in the r\'egime of cosmological perturbations.

As in other modified gravity theories, the two metric perturbations, the Newtonian potential $\Phi$ and the metric potential $\Psi$, evolve differently even with no anisotropic stress. Contrarily, for general relativity $\Phi=-\Psi$ holds in the matter dominated era. Thus, when we study gravitational lensing we have to deal with the deflecting potential $\Upsilon\equiv(\Psi-\Phi)/2$. Moreover, its Poisson equation, which relates it to the distribution of the matter overdensities, is modified by the presence of an effective time- and scale-dependent gravitational constant.

It is worth giving a final remark on the evolution of matter fluctuations. Unlike the linear growth of perturbations, that can be described analytically, the non-linear r\'egime has to be explored numerically. Two approaches have been followed and we refer to them as ``KW'' and ``PPF.'' The former generalises the \textsc{halofit} procedure \citep{Smith:2002dz} to the eDGP scenario according to recently performed $N$-body simulations \citep{Khoury:2009tk}, whilst the latter interpolates the eDGP non-linear matter power spectrum with that of \lcdm\ in order to reproduce general relativity at small scales and be thus able to pass Solar system gravity tests \citep{Hu:2007pj}. The functional forms of this last approach have been obtained by perturbation theory \citep{Koyama:2009me} and confirmed by $N$-body simulations \citep{Schmidt:2009sg}. Unfortunately, were PPF the correct non-linear prescription, we would not be able to discriminate between the eDGP and \lcdm\ signals even with the present and next generation weak lensing surveys \citep{Camera:2011mg}.

Finally, we analyse the St and HS $f(R)$ theories, which also are degenerate with \lcdm\ at background level \citep{Cardone:2010}. Their functional form allow them to achieve the late-time accelerated expansion of the Universe with no formal cosmological constant. On the contrary, they present three free parameters, $c_1$, $c_2$ and $n$. It has been shown that the growth of linear perturbations strongly depends on the function $f(R)$, which acts by generating a time- and scale-dependent gravitational constant, as well as an effective anisotropic stress \citep{Tsujikawa:2007gd}. Regarding the non-linear evolution of perturbations, the PPF technique is still valid, as confirmed by $N$-boy simulations \citep{Oyaizu:2008tb}. Cosmic shear studies on these models gave interesting results \citep{Beynon:2009yd,Thomas:2011pj}, but the St signal is nonetheless almost completely degenerate with \lcdm\ \citep{Camera:2011ms}.

\textit{Results and Discussion.---} Here, we present the cosmic flexion power spectrum \eqref{eq:C^F_l} expected in the alternative models outlined above and we compare it with the \lcdm\ prediction. For this, we use a fiducial flat Universe where the Hubble constant is $\ho=100\,h\,\mathrm{km~s^{-1}~Mpc^{-1}}$ and $h=0.7$. The matter density in units of the critical density is $\om\equiv\odm+\ob=0.28$, with $\odm$ and $\ob=2.22\cdot10^{-2}h^{-2}$ the dark matter and baryon fractions, respectively. The tilt of the primordial matter power spectrum is $n_s=0.96$ and the density fluctuation rms on the scale of $8\,h^{-1}\,\mathrm{Mpc}$ is $\sigma_8=0.8$. For the UDM model, we probe $\cinf=5\cdot10^{-4}$ and $\cinf=10^{-3}$. The eDGP model parameters are $\alpha=0.116$ and $\rc\ho=155.041$ \citep{Camera:2011mg}. The St(HS) parameters read $\log_{10}c_1=2.38(4.98)$, $\log_{10}c_2=-2.6(3.79)$ and $n=1.79(1.67)$ \citep{Cardone:2010}.

It is important to note that there currently is no linear-to-non-linear mapping in UDM models. Nevertheless, differences between \lcdm\ and UDM models arise at scales smaller than the sound horizon. With a cross-over wavenumber $k\simeq1/\lambda_J$, if the sound speed is small enough to guarantee that $\lambda_J$ is well within the non-linear regime, we can assume that the non-linear evolution of the UDM power spectrum will be similar to that of \lcdm\ \citep{Camera:2010wm}.

We use the specifics of the upcoming ESA Euclid satellite \citep{Refregier:2010ss}.\footnote{\texttt{http://sci.esa.int/science-e/www/area/index.cfm?fareaid=102}} Euclid is one of the ESA Cosmic Vision 2010-2015 approved projects and is currently in the timeline of M-class missions. Its survey area will be $20,000$ square degree, with a sky coverage $f_\mathrm{sky}\simeq0.48$ and a source distribution over redshifts \citep{1994MNRAS.270..245S}
\begin{equation}
n(z)\propto z^2e^{-\left(\frac{z}{z_0}\right)^{1.5}},\label{eq:n_z-Euclid}
\end{equation}
where $z_0=z_m/1.4$ and $z_m=0.9$ is the median redshift of the survey. The number density of the sources, with redshift and shape estimates, is $\bar n=35\,\mathrm{arcmin^{-2}}$. To compute errorbars, we use
\begin{equation}
\Delta C^{\mathcal F}(\ell)=\sqrt{\frac{2}{(2\ell+1)f_\mathrm{sky}}}\left[C^{\mathcal F}(\ell)+N^{\mathcal F}_\ell\right],\label{eq:errorbars}
\end{equation}
generalising thus the approach of \citep{Kaiser:1991qi}. This is because -- unlike shear -- flexion has a dimension of $\mathrm{length}^{-1}$ (or $\mathrm{angle}^{-1}$). This means that the effect by flexion depends on the source size. Recently, it has been shown that the noise power spectrum $N^{\mathcal F}_\ell$ for flexion is inversely proportional to the squared angular scale \citep{Pires:2010ar}; we therefore set
\begin{equation}
N^{\mathcal F}_\ell=\frac{4\pi^{2}\langle{{\mathcal F}_\mathrm{int}}^2\rangle}{\ell^{2}\bar n},
\end{equation}
with $\langle{{\mathcal F}_\mathrm{int}}^2\rangle^{0.5}\simeq0.03\,\mathrm{arcsec^{-1}}$ the galaxy-intrinsic flexion rms.

\begin{figure}[!h]
\centering
\includegraphics[width=0.5\textwidth]{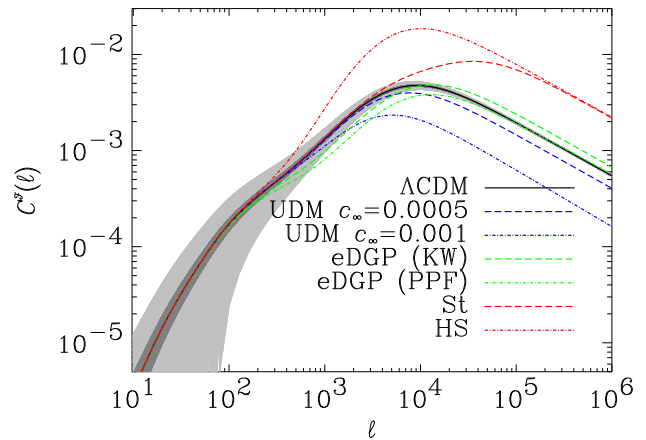}
\caption{Power spectrum of cosmic flexion $C^\mathcal F(\ell)$ for the alternative cosmological models presented in the text.}\label{fig:C^FF}
\end{figure}
Fig.~\ref{fig:C^FF} shows the cosmic flexion power spectra $C^{\mathcal F}(\ell)$ of \lcdm\ (red, solid), eDGP (green) with both KW (dashed) and PPF (dot-dashed) linear-to-non-linear mappings and the $f(R)$ models (red) of St (dashed) and HS (dot-dashed). As expected, the UDM signal is suppressed at small angular scales because of the presence of the scalar field sound speed. The eDGP model is still very close to the \lcdm\ prediction, particularly the PPF non-linear power spectrum, for it being specifically designed to reproduce general relativity on small scales. On the other hand, the St and HS models clearly show the scale dependence of the Newtonian gravitational constant $G$. Indeed, in the so-called ``scalar-tensor'' r\'egime it reaches the value $\sim4G/3$ \citep{Tsujikawa:2007gd}.

Nevertheless, cosmic flexion shows an outstanding improvement in the separation between the signals when compared to the cosmic shear power spectrum. Indeed, the dark-grey shaded area represents the $1\sigma$-error region, whilst light-grey refers to errors \textit{six times} larger. Flexion measurements are made on the shapes of the source galaxies, exactly as in the cosmic shear analysis. Therefore, the source number density $\bar n$ is the same for the two observables and, with a space-based, wide-field survey such as Euclid, we can collect a fairly large statistics. However, the intrinsic flexion rms $\langle{{\mathcal F}_\mathrm{int}}^2\rangle^{0.5}$ is an order of magnitude smaller than the cosmic shear rms and the power spectrum is thus significantly less noisy.

We conclude that cosmic flexion is an excellent tool for testing alternative cosmological models and discriminate between them. With  realistic values for the mean galaxy number density $\bar n$ and the flexion noise $N^{\mathcal F}_\ell$, which includes its angular scale-dependence  \citep{Pires:2010ar}, expected from the upcoming Euclid mission, we find an admirable separation between cosmic flexion power spectra $C^{\mathcal F}(\ell)$ of viable models which are almost degenerate with \lcdm\ when investigated with other observables, such as cosmic shear. We will provide a more detailed analysis of these outstanding results in \citep{Camera:flexion}.

SC and AD acknowledge support from the INFN grant PD51 and the PRIN-MIUR-2008 grant 2008NR3EBK ``Matter-antimatter asymmetry, dark matter and dark energy in the LHC era.'' This research has made use of NASA's Astrophysics Data System.

\bibliographystyle{apsrev4-1}
\bibliography{/home/camera/Documenti/LaTeX/Bibliography}

\end{document}